\documentclass[iop]{emulateapj-rtx4} % read by: ACROREAD
\shortauthors{Sekanina}
\shorttitle{Inevitable Endgame of Comet C/2023 A3}
\slugcomment{Version \today }

\begin{document}
\title{Inevitable Endgame of Comet Tsuchinshan-ATLAS (C/2023 A3)\\[-1.5cm]}
\author{Zdenek Sekanina}
\affil{La Canada Flintridge, California 91011, U.S.A.; {\sl ZdenSek@gmail.com}}
% \email{ZdenSek@gmail.com.}
\begin{abstract} % maximum length = 1920 characters
Hopes are being widely expressed that C/2023 A3 could become a naked-eye
object about~the~time~of its perihelion passage in late 2024.  However,
based on its past and current performance, the~comet~is expected to
disintegrate before reaching perihelion.  Independent lines of evidence
point to its forthcoming inevitable collapse.  The first issue, which was
recently called attention to by I.~Ferrin, is this Oort cloud comet's
failure to brighten at a heliocentric distance exceeding 2~AU, about
160~days~pre\-perihelion, accompanied by a sharp drop in the production
of dust ({\it Af}$\rho$).  Apparent over a longer~period of time, but
largely ignored, has been the barycentric original semimajor axis inching
toward~negative numbers and the mean residual increasing after the light-curve
anomaly, suggesting a fragmented nucleus whose motion is being affected
by a nongravitational acceleration; and an unusually narrow, teardrop dust
tail with its peculiar orientation, implying copious emission of large
grains far from the Sun but no microscopic dust recently.  This evidence
suggests that the comet has~entered~an~advanced phase of fragmentation,
in which increasing numbers of dry, fractured refractory solids
stay~assembled in dark, porous blobs of exotic shape, becoming undetectable
as they gradually disperse~in~space.
\end{abstract}

\section{Introduction} % Sec. 1
Initially discovered at the Purple Mountain Observatory's XuYi Station,
China, on 9~January 2023, then lost and discovered a second time at the {\it
Asteroid Terrestrial-Impact Last Alert System\/} (ATLAS) search project's
station at Sutherland, South Africa, 44~days later (Green 2023),
comet C/2023~A3 has been monitored for nearly 17~months at the
time of this writing and it still has more than two months to go before
reaching perihelion.  Since its minimum distance from the Sun is to be
merely 0.39~AU, the comet has been widely predicted to become a very
bright, possibly naked-eye object around perihelion, especially in
early October, in part because of effects of forward scattering of
sunlight by microscopic dust at phase angles near 180$^\circ\!$.

In the past, such predictions for other comets with similar parameters
(discovered long before perihelion as a fairly bright object, approaching
the Sun to a small fraction of the Sun-Earth distance) failed miserably
on a number of occasions, but proved correct in some cases.  Performance
prognostication for individual objects is a high-risk branch of cometary
science, which is very slowly being mastered in the process of our gradually
learning to understand the whims of comets.  Yet, it appears that at least
some comets do signal early messages that correlate with their subsequent
behavior.  Timely recognition and proper interpretation of these messages
should pave the way for a more successful future in this unusual field of
scientific endeavor.

Comet C/2023 A3 appears to be a member of the class of undersized Oort
cloud objects in orbits with small perihelion distances, many of which
turned out to be unable to cope with the physical conditions they were
subjected to in the inner Solar System and disintegrated as a result of
progressing fragmentation.  In the following I address the traits of
comet C/2023 A3 that appear to be of the same or similar nature.

\section{The Light Curve, Dust Content, and\\Production of Water} 
With all pre-discovery images the comet's light-curve database covers
about 27~months.  The dependence of the total brightness, normalized
to a unit geocentric distance, $\Im_\Delta$, on the heliocentric
distance,~$r$, has often been expressed by a power law,
\mbox{$\Im_\Delta = \Im_0 \, r^{-n}$}, whose parameters are the
absolute magnitude \mbox{$H_0 = -2.5 \log \Im_0$} and exponent $n$,
which determines the {\vspace{-0.035cm}}slope $d\log \Im_\Delta/d\log
r$.\footnote{A comet's brightness could also be corrected for
a phase effect, if phase angles cover a wide range or are near
180$^\circ\!$.} For \mbox{example}, S.~Yoshida\footnote{See {\tt
http://www.aerith.net/comet/catalog/search.cgi}.} has used \mbox{$H_0
= 4.5$} and \mbox{$n = 4$} to fit all magnitude observations
made between 2~March 2023 and 24~January 2024.  Similarly,
A.~Kammerer\footnote{See {\tt https://fg-kometen.vdsastro.de}.}
has found  \mbox{$H_0 = 4.7$} and \mbox{$n = 3.96$} from 384~data
reported by 64~observers until the end of April 2024.

Because of time constraints, I have not investigated the entire light
curve of the comet.  From the standpoint of predicting the comet's
evolution in the near future, it is the recent activity that is
most critical.  To assess it, I deemed it sufficient to inspect
a representative light curve based on the data reported by a single
observer, but on the condition of their maximum possible homogeneity
to mitigate a contamination as much as possible.

From the {\it Comet Observation Database\/}\footnote{See {\tt
https://cobs.si}.} (COBS) website I eventually selected a set of
48~unfiltered CCD total-magnitude observations made by A.\,Pearce
with a 35-cm f/5 Schmidt Cassegrain between 21~January and 13~June
2024.  Although the contribution from the outer coma may be
unaccounted for, this is not a major problem, as the emphasis is
on rather abrupt brightness variations that begin at the nuclear
condensation.\footnote{I ignored the phase effect, because
in the investigated period of time the phase angle varied only
between 3$^\circ$ and 28$^\circ\!$.}

\begin{figure}[t] % Figure 1
\vspace{0.17cm}
\hspace{-0.2cm}
\centerline{
\scalebox{0.565}{
\includegraphics{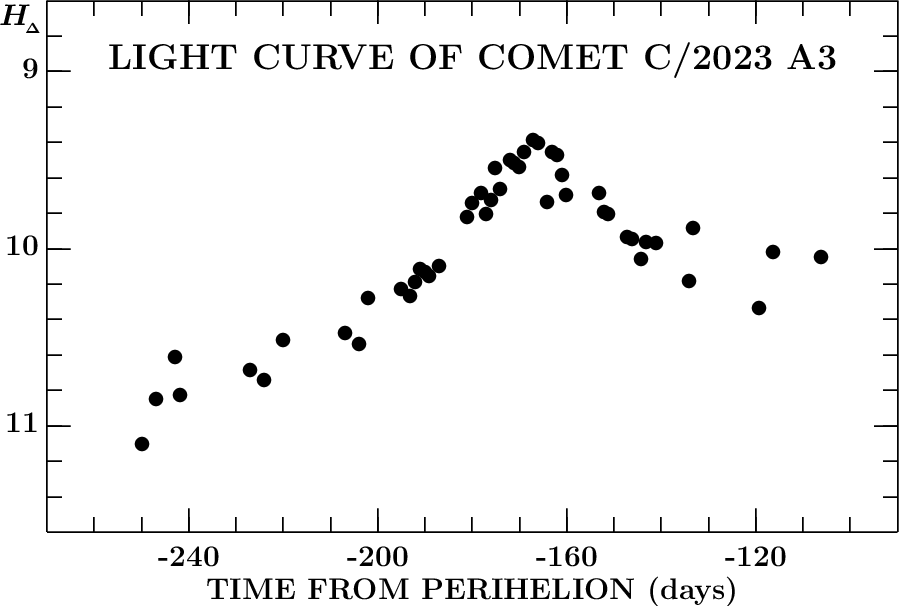}}}
\vspace{-0.05cm}
\caption{Light curve of comet Tsuchinshan-ATLAS between 250 and
106~days before perihelion (21~January through 13~June 2024), as
portrayed by a set of 48 CCD total-magnitude observations made
by A.\,Pearce with his 35-cm f/5 Schmidt-Cassegrain.  The ordinate
is the total observed CCD magnitude of the comet normalized to a
unit geocentric distance, $H_\Delta$.  Surprisingly, the comet was
brightening at an accelerating rate before 165~days preperihelion
(15~April 2024), but began to fade afterwards.{\vspace{0.6cm}}}
\end{figure}

To obtain an insight into the comet's activity variations over the past
five months covered by Pearce's observations, I plotted the magnitudes,
normalized to 1~AU from the Earth, $H_\Delta$, as a function of time.
This plot, displayed in Figure~1, has immediately raised the red flag
because it shows a dramatic change at 165~days before perihelion,
on 15~April 2024, when the intrinsic brightness reached a peak:\
the comet was brightening at a rate that appears to have been
increasing with time before this date, but began to fade afterwards.
This finding supports recent concerns expressed by I.~Ferrin\footnote{See
{\tt https://groups.io/g/comets-ml/messages?dir=desc}.} (e.g., his
message \#32307).

To further investigate the comet's activity, I examined the same
dataset on a plot that displays the normalized magnitude $H_\Delta$
as a function of heliocentric distance $r$ (on a logarithmic scale).
This plot, presenred in Figure~2, suggests that the early segment of the
light curve, before the brightness peaked, may have consisted of two parts.
Up to 190~days before perihelion (21~March 2024), when the comet was
3.3~AU from the Sun, its light curve was described by the absolute
magnitude \mbox{$H_0 = 5.6 \pm 0.4$} and the slope parameter
\mbox{$n = 3.5 \pm 0.3$}.  Hence, the absolute brightness was only about
1~mag fainter than given by both Yoshida and Kammerer, as mentioned
above, while the slope was slightly lower than the results of the two
authors indicated.  This suggests that up to 190~days before perihelion
the comet showed only a moderate rate of progressive fading compared
to earlier times.

\begin{figure}[t] % Figure 2
\vspace{0.15cm}
\hspace{-0.19cm}
\centerline{
\scalebox{0.565}{
\includegraphics{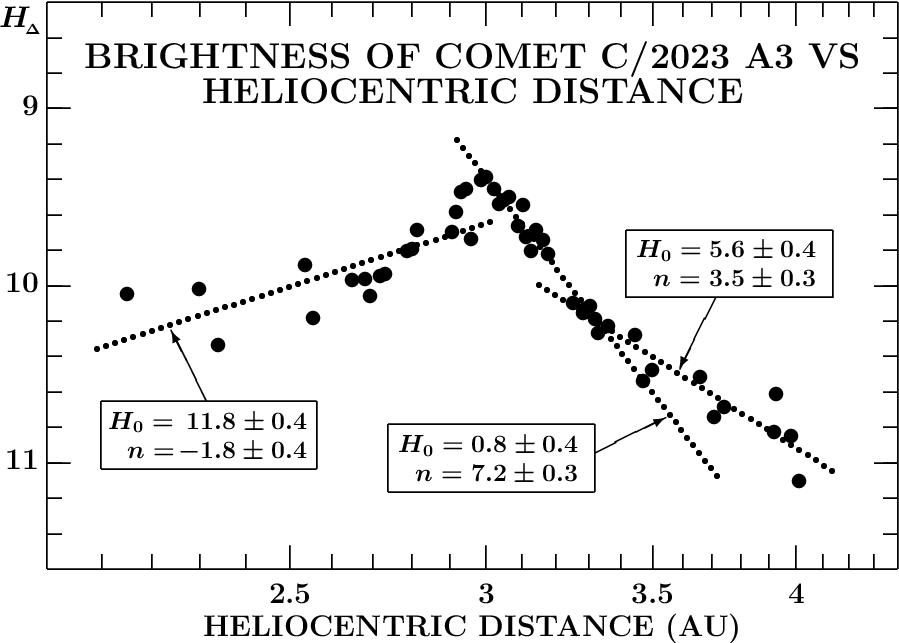}}}
\vspace{-0.05cm}
\caption{Intrinsic brightness of comet Tsuchinshan-ATLAS as a
function of heliocentric distance.  The set of magnitude data~from
Figure~1 now shows three likely stages of evolution, as
the comet's heliocentric distance was decreasing from 4~AU
to 2~AU.  At $>$3.3~AU the light curve followed a law that
did not differ too much from the law that applied far from
the Sun.  Between 3.3~AU and 3~AU the comet underwent a surge
of activity, caused apparently by prolific fragmentation of the
nucleus, which resulted in doubling the value of the slope
parameter $n$ and a jump in the absolute brightness by
5~magnitudes.  By the time the comet reached 3~AU the
out-of-control fragmentation of the nucleus ceased on the
global scale, with further fracturing proceeding locally.  This
trend is seen to have continued for more than two months and
is expected to continue until the ultimate collapse of the
comet still before it reaches perihelion.{\vspace{0.4cm}}}
\end{figure}

The latter part of the period of brightening in Figures~1 and 2,
starting about 190~days before perihelion and 3.3~AU from the Sun,
shows a rapid reactivation of the comet.  One cannot call it an
outburst because the rate of reactivation was not steep enough.
However, compared to the previous part of the light curve, $n$
doubled to \mbox{$n = 7.2 \pm 0.3$} and $H_0$ shot up by about
5~magnitudes to \mbox{$H_0 = 0.8 \pm 0.4$}.  The episode could
perhaps be described as a surge of activity.  It is proposed that,
if not earlier, it was about 21~March that the nucleus began to
fragment profusely.  Emission from a rapidly expanding surface area
of short-lived activity could explain the stepped-up brightening,
which continued over a period of 25~days (from 21~March to 15~April
2024) at heliocentric distances from 3.3 to 3.0~AU, ending up with
the peak at 165~days before perihelion.  Once the source of extra
emission got exhausted, the remaining mass began to underperform,
resulting in the light curve's downturn.  Occasional flare-ups caused
by further episodes of local fragmentation are likely to be the cause
of large scatter among the data after 15~April.  This scenario has now
continued for more than two months and is expected to continue until
the ultimate disintegration of the nucleus.  The downward trend is
described by somewhat uncertain parameters:\ \mbox{$H_0 = 11.8 \pm 0.4$}
and \mbox{$n = -1.8 \pm 0.4$}.

The evidence from the light curve is strongly supported by the data
on a dust-content parameter {\it Af}$\rho$, introduced by A'Hearn
et al.\ (1984) as a proxy to{\vspace{-0.04cm}} estimate the dust
production rate.  A website by {\it Cazadores de Cometas}$\:$\footnote{See
{\tt http://astrosurf.com/cometas-obs}.} (Comet Hunters), which
contains, among other information, an {\it Af}$\rho$ database by
mostly Spanish observers (including the Canary Islands), offers
more than 150~data points for comet Tsuchinshan-ATLAS, starting
in late February 2023.  One of the plots shows that, reduced to a
coma 10\,000~km in radius, {\it Af}$\rho$ gradually increased from a
minimum of $\sim$500~cm in early March to $\sim$1800~cm in mid-August
2023.  Between December 2023 and February 2024 {\it Af}$\rho$ stayed
nearly constant at 4000~cm, but in March it started increasing
sharply, reaching a peak of $\sim$9000~cm by mid-April, followed
by a steep drop.

\begin{figure} % Figure 3
\vspace{0.17cm}
\hspace{-0.18cm}
\centerline{
\scalebox{0.52}{
\includegraphics{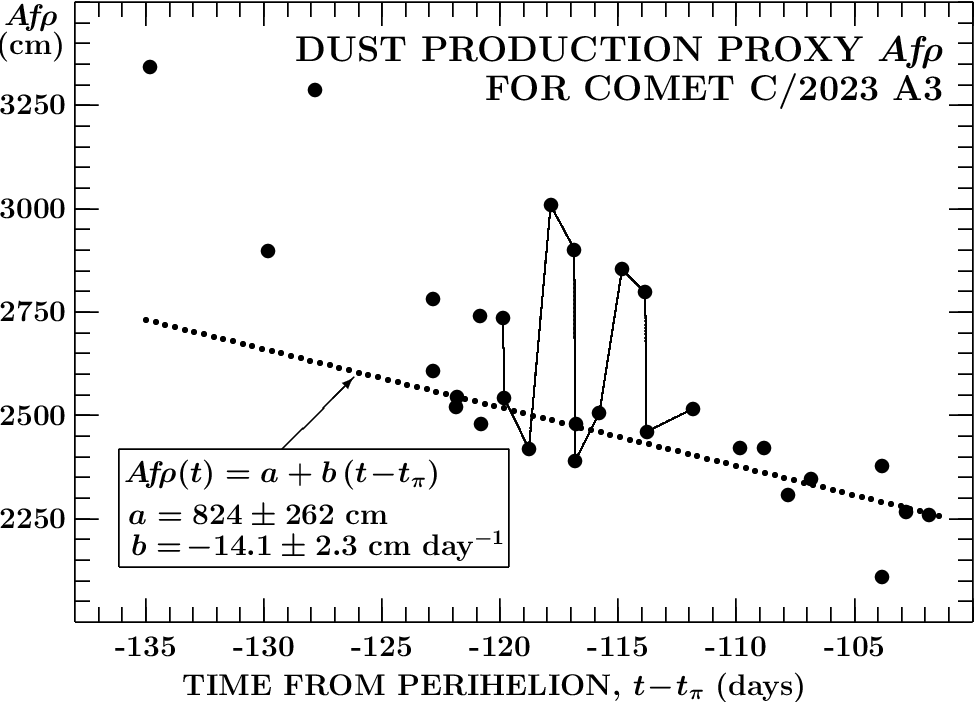}}}
\vspace{-0.1cm}
\caption{Variations in the dust production parameter {\it Af}$\rho$
of comet Tsuchinshan-ATLAS between 15~May and 17~June 2024 (135 to
102~days before perihelion).  The large scatter is not necessarily
caused in its entirety by errors of observation, but could in part
be a product of sudden dust release during episodes of fragmentation,
as depicted for the two major features between 120 and 112~days
before perihelion.  The {\it Af}$\rho$ data points near the bottom
appear to decrease systematically with time at a crude average rate
of $\sim$14~cm per day.{\vspace{0.6cm}}}   
\end{figure}

The curve of {\it Af}$\rho$ fairly closely parallels the light
curve.  Particularly interesting are the variations between mid-May
and mid-June (from 135 to 102~days before perihelion), which are
reproduced in Figure~3.  The considerable scatter, similar to the
scatter in the light curve, is believed to be in part triggered by
short-term release of dust during minor episodes of nuclear fragmentation.
The rest of the scatter could be errors of observation.  The baseline
data in Figure~3 are shown to follow an approximately linear
systematic decrease with time at a rate of about 14~cm day$^{-1}$,
with an extrapolated value of more than 800~cm at perihelion.

It has been known for a fairly long time that some faint comets
in orbits with small perihelion distance perish shortly before
perihelion (Bortle 1991), although Jewitt \& Luu (2019) did
recently report a comet that disintegrated at 1.9~AU from the
Sun.  My investigation of the problem led to a conclusion that
the objects that perish are always intrinsically faint Oort
cloud comets depleted in dust (Sekanina 2019).  To be able to
predict the chance of survival from early observations, I
introduced a {\it synoptic index for perihelion survival\/},
$\Im_{\rm surv}$, given by a formula
\begin{equation}
\Im_{\rm surv} = H_0 - 5.7 - \frac{35}{6} \log q - \frac{5}{3}
 \log (\!A\!f\!\rho)_0,
\end{equation}
where $H_0$ is the absolute magnitude, $q$ is the perihelion distance
in AU, and ($\!${\it Af}$\rho)_0$ (in cm) is strictly to be taken at 1~AU
from the Sun and at a zero phase angle.  A comet is predicted to
survive perihelion when \mbox{$\Im_{\rm surv} < 0$} and vice versa.

In the following section I will demonstrate that comet Tsuchinshan-ATLAS
unquestionably has arrived from the Oort cloud, so that application
of the synoptic index is relevant.  Even though the quantities that
determine the index are rather uncertain at present, it nonetheless is
of inrerest to compute a preliminary value of $\Im_{\rm surv}$.  To do
so, I first of all adopt \mbox{$H_0 = 11.8$} from the review of the
light curve in Figure~2.  Next, from its dependence on time proposed
in Figure~3 I compute {\it Af}$\rho$ at time~$t_0$ when the comet
reaches 1~AU from the Sun preperihelion, \mbox{$t_0 \!-\! t_\pi =
-38$ days}.  The result is \mbox{($\!${\it Af}$\rho)_0 = 1360$ cm}.  With
the perihelion distance of 0.391~AU, the synoptic index amounts to
\begin{equation}
\Im_{\rm surv} = +3.3
\end{equation}
and the comet is predicted to have no chance of surviving perihelion.

Of course, one may question whether the linear extrapolation of
{\it Af}$\rho$ is appropriate and the employed values of $H_0$ and
($\!${\it Af}$\rho)_0$ compatible.  As a check, I fit the relevant
data from Figure~3 by a power law of heliocentric distance $r$,
similar to the light curve:
\begin{equation}
A\!f\!\rho(r) = (\!A\!f\!\rho)_0 \: r^{-z}
\end{equation}
and obtain by least squares \mbox{($\!${\it Af}$\rho)_0 = 1170
\pm 40$ cm} and \mbox{$z = -0.90 \pm 0.15$}.  The synoptic index
now equals
\begin{equation}
\Im_{\rm surv} = +3.4,
\end{equation}
very close to the result (2).

In summary, from the standpoint of the perihelion survival of comet
Tsuchinshan-ATLAS, both its systematic fading documented by the light
curve between mid-April and mid-June 2024 and the continuing drop of
{\it Af}$\rho$ are extremely worrisome.   The synoptic index predicts
that the comet will perish before perihelion, as an instrumental
correction of $H_0$ could not possibly exceed 1--2 mags.\,

Independent of this conclusion, I should note that from
the spectra taken on 31~May 2024 when the comet was 2.33~AU from
the Sun, Ahuja et al.\ (2024) obtained two major results.  One, the
production rate ratio $Q$(C$_2$)/$Q$(CN)\,$<$\,0.32, suggesting a
{\vspace{-0.04cm}}carbon depleted comet; and two, \mbox{$Q$(H$_2$O)\,=\,$(1.50
\pm 0.37) \!\times\! 10^{28}$\,s$^{-1}$}, an unexpectedly high
rate of water production.  At 2.33~AU from the Sun, a spatially
averaged sublimation rate of water ice is \mbox{$1 \!\times\!
10^{17}$\,cm$^{-2}$\,s$^{-1}$}, suggesting that the total sublimating
surface was 15~km$^2$!  As only a small fraction of a comet's surface
is known to be active, a monolithic nucleus would have been close to
10~km across --- an absurd result.   An unresolved cloud of sublimating
boulder-sized fragments offers a more plausible solution.\,\,\,

\begin{table*}[t] % Table 1
\vspace{0.19cm}
\hspace{-0.2cm}
\centerline{
\scalebox{1}{
\includegraphics{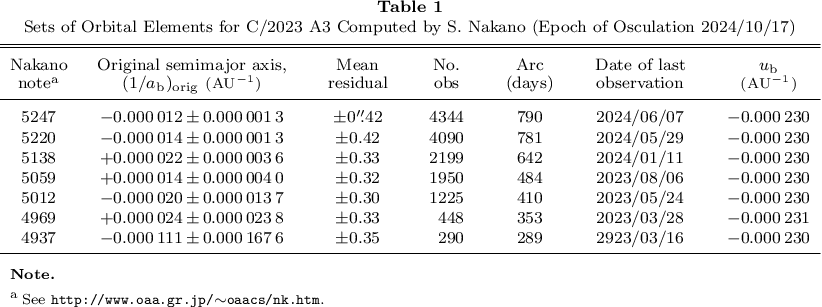}}}
\vspace{0.75cm}
\end{table*}

Consider $\nu$ fragments of an equal volume.  Let the effective diameter
of the parent nucleus before a fraction $\chi$ of its volume fragmented
be $D_{\rm par}$ and the effective diameter of each fragment $D_{\rm frg}$.
Then \\[-0.15cm]
\begin{equation}
\chi D_{\rm par}^3 = \nu D_{\rm frg}^3. \\[-0.05cm]
\end{equation}
Let the sublimating fraction of the surface of the parent nucleus
be \mbox{$f_{\rm par} \ll 1$}.  Since the previously hidden~ice-rich
surface is, because of fragmentation, being exposed to sunlight for
the first time, one can assume that for an overwhelming majority of
fragments their entire surface facing the Sun is sublimating, so that
\mbox{$f_{\rm frg} = \frac{1}{2}$}.  The active surface area before
fragmentation is{\vspace{-0.05cm}}
\begin{equation}
S_{\rm par} = \pi f_{\rm par} D_{\rm par}^2, \\[-0.05cm]
\end{equation}
but after fragmentation it is \\[-0.15cm]
\begin{equation}
S_{\rm frg} = {\textstyle \frac{1}{2}} \pi \nu D_{\rm frg}^2
 = {\textstyle \frac{1}{2}} \pi \nu^{\frac{1}{3}} \!\chi^{\frac{2}{3}}\!
 D_{\rm par}^2, \\[-0.05cm]
\end{equation}
{\nopagebreak}where I inserted for $D_{\rm frg}$ from Equation~(5).
Here $S_{\rm frg}$ is{\pagebreak} known from the observation.  The
ratio of $S_{\rm frg}/S_{\rm par}$ is
\begin{equation}
\frac{S_{\rm frg}}{S_{\rm par}} = \frac{\nu^{\frac{1}{3}}
 \chi^{\frac{2}{3}}}{2f_{\rm par}} .
\end{equation}
From Equation (7) the number of fragments needed to satisfy the observed
large active surface $S_{\rm frg}$ equals
\begin{equation}
\nu = \!\left( \!\frac{2 S_{\rm frg}}{\pi} \!\right)^{\!\!3} \!
 D_{\rm par}^{-6} \, \chi^{-2}\!,
\end{equation}
and the effective diameter of a fragment is
\begin{equation}
D_{\rm frg} = \frac{\pi D_{\rm par}^3 \, \chi}{2 S_{\rm frg}}.
\end{equation}
The counterintuitive inverse correlation between $\nu$ and $\chi$ is
explained by the fragment size varying as $\chi$.

It is enlightening to see the results after inserting some plausible
numbers.  The sublimation constraints dictate that \mbox{$S_{\rm
frg} = 15$ km$^2$}.  From the orbital considerations in the next
section one should rather prefer that $D_{\rm par}$ be below 1~km.
Then, for example, fracturing 75~percent of the volume of a
nucleus 0.5~km in diameter will produce $\sim$100\,000~fragments,
each 10~meters across.  In reality, the fragments have of course
some size distribution.  The active-surface ratio $S_{\rm frg}/S_{\rm
par}$ {\vspace{-0.05cm}}becomes 190, if \mbox{$f_{\rm par} = 0.1$},
so that $S_{\rm par}$ is less than 0.1~km$^2$.

\section{The Orbit}
\mbox{In nearly$\:$17$\;$months that have elapsed since its second}
discovery, the comet was observed for astrometry more than 4000 times.
In addition, almost 30 pre-discovery images were subsequently
located, obtained between 9~April and 22~December~2022 at three
observatories:\ Panoramic Survey Telescope and Rapid Response~System
(Pan-STARRS 2), Haleakala Observatory (code F52); Dark
Energy Camera Project (DECam), Cerro Tololo Inter-American
Observatory (code W84); and Zwicky Transient Facility (ZTF),
Palomar Observatory (code I41).  As a result, at the time of
this writing the observed arc of the orbit equals close to
27~months.  One would expect that the orbital elements,
including the semimajor axis, should by now be determined with
very high accuracy.  As described in some detail in the following,
inspection of the published computation results suggests that this
indeed is the case, even though there is a hitch.

I start with the work by S.\ Nakano, who has been improving the
orbital elements of this comet on a number of occasions.  The
selected data from his results, relevant to the issues of interest
here, are summarized in Table~1.  Presented in the individual
columns are the reference; the reciprocal barycentric original
semi-major axis, $(1/a_{\rm b})_{\rm orig}$ in {\small AU$^{-1}$}
(the most important column); the mean residual; the number of
astrometric observations used; the orbital arc covered by
these observations (in days); the date of the last observation;
and the difference between the reciprocal heliocentric osculating
semi-major axis (whose value depends on the choice of the epoch
of osculation) and the quantity in column~2, \mbox{$u_{\rm b} =
(1/a)_{\rm osc} \!-\! (1/a_{\rm b})_{\rm orig}$}.  Since Nakano
has always been using the standard epoch of osculation, $u_{\rm b}$
should essentially be constant.

The tabulation of $u_{\rm b}$ is intended to support the proposed
explanation of a discrepancy with the results obtained
at the {\it Minor Planet Center\/} (MPC), which are with two
additional independent results listed in Table~2, whose format
is similar to that of Table~1.  Comparison of the orbital solutions
that include recent observations shows an excellent agreement between
the values of $(1/a_{\rm b})_{\rm orig}$ by JPL and by Williams,
their{\vspace{-0.04cm} difference amounting to less than 1~unit of
10$^{-6}$\,AU$^{-1}$, within the errors of observation.  Next
comes Nakano's most recent solution, which gives a value that is
more negative by 11--12~units of 10$^{-6}$\,AU$^{-1}$.  This is
not surprising, as Kr\'olikowska \& Dones' (2023), who examined
differences among the various methods of orbit determination, have
found that Nakano's values of $(1/a_{\rm b})_{\rm orig}$ have
systematically been lower by about 10~units.  Whatever the source
of this discrepancy, once it is accounted for, Nakano's result agrees
with the other two to 1--2~units as well.

\begin{table*}[t] % Table 2
\vspace{0.19cm}
\hspace{-0.2cm}
\centerline{
\scalebox{1}{
\includegraphics{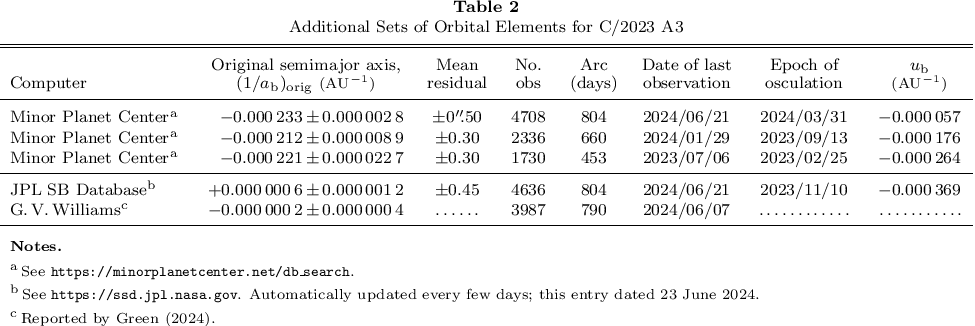}}}
\vspace{0.45cm}
\end{table*}

The huge problem appears to be with the MPC orbits, whose values of
$(1/a_{\rm b})_{\rm orig}$ exhibit major deviations.  The discrepancy
is so large that its source must be a major oversight.  I note that
the first MPC value of $(1/a_{\rm b})_{\rm orig}$, in particular,
is very close to Nakano's value of $u_{\rm b}$.  The possible
confusion between these two quantities could explain the problem.
On the other hand, this just may happen to be a coincidence,
because the respective epochs of osculation are more than six
months apart and $u_{\rm b}$ may have changed substantially.
While the source of the discrepancy remains unclear, I accept 
that the barycentric original orbit of the comet based on the
observations available at present is essentially a parabola.

The orbit determination of comet Tsuchinshan-ATLAS involves more
subtle issues related to the question of the object's survival.
For example, no nongravitational orbital solution has as yet been
published, apparently because the gravitational solutions have
been entirely satisfactory.  And this may have been so, at least
in part, because the comet has so far been relatively far from
the Earth and the amplitude of potential systematic residuals
small.  This will of course change in the future and so will
other things.

One point of concern that I see from Nakano's results is a recent
trend to more negative values of $(1/a_{\rm b})_{\rm orig}$.  The
change is about 30 units or more than 7$\sigma$ of the error in the
5059 run and more than 20$\sigma$ of the error in the 5247 run.  If
this trend should continue, it could represent an independent
confirmation of progressive fragmentation of the nucleus.  This
trend would necessitate to eventually incorporate the nongravitational
terms.

I call attention to this effect for two reasons.  One is that the
trend toward more negative values appears in the first orbital
solutions that incorporate the observations made following the
light-curve anomaly in mid-April.  The other reason is that the
mean residual from these solutions has increased relative to the
previous solutions by 0$^{\prime\prime\!}$.1 or about 30~percent.
That is a lot and it indicates that for some reason the nuclear
condensation has been measured with lesser accuracy.  I will
return to this issue and its implications in the following section.

The observers in the south have at most until the end of July to
monitor the comet before it disappears in the solar glare; it will
then still be more than 1~AU from the Sun.  It should be seen no
more afterwards.

\section{The Tail}
The tail completes the list of peculiarities that are being displayed
by this unusual object.  Plasma tails are seldom detected farther than
2~AU from the Sun, so it is not surprising that the tail of comet
Tsuchinshan-ATLAS is definitely composed of dust.  Except on special
occasions that I address later, preperihelion dust tails are --- because
of dynamical constraints --- rather narrow, but seldom as narrow and
teardrop-shaped as displayed by this object.  They typically exhibit
two characteristics:\ have a tendency, however slight, to broaden with
increasing distance from the comet's head and extend to a maximum
distance along an axis that makes a small angle with the prolonged
radius vector.

The tail of comet Tsuchinshan-ATLAS defies either of the two rules.
Fortunately, this controversial behavior provides interesting information
about the comet's history and activity.  Tails of this appearance and
outlines, especially the tendency of getting narrower farther from the
head, have in the past been observed primarily among comets arriving
from the Oort cloud in orbits with large perihelion distances.  The
tails of these comets deviated considerably from the radius vector and
their investigation had a rather bizarre history (e.g., Osterbrock 1958,
Brandt 1961, Roemer 1962, Belton 1965, Sekanina 1973, Meech et al.\ 2009).

To describe the physical significance of the odd-shaped tail of comet
Tsuchinshan-ATLAS, I have selected one of many available images (which
all look alike) from the recent past and I am going to explain the
meaning of the displayed features below.

The selected picture, in Figure 4, has been cropped from a false-color
computer-processed rendition of the image taken by R.\ Naves, of the
Observatorio Montcabre, with a 30-cm f/9 reflector on 4 June 2024.  It
is a long, nearly two-hour exposure, which is important in that it
shows the true extent, in excess of 8$^\prime$, of the very delicate
tail.  The price paid is the apparent saturation of the central coma,
whose projected radius of 13$^{\prime\prime}$ is equivalent to
17\,000~km.

Figure 4 consists of two parts:\ the upper panel displays the
image, while the lower panel offers a guide to its interpretation.
Detailed information provided by Naves on the details of his
observation, including the orientation and scale, has been most
helpful for an accurate description of the properties of the tail.
Besides its getting more narrow with increasing distance from the
head, a rather stunning trait is its large angular separation of
16$^\circ$ from the projected radius vector, the maximum dynamically
allowed being 22$^\circ\!$.5 but plausible only 19$^\circ$.  More
than 100~days preperihelion in an orbit that comes to less than
0.4~AU from the Sun, this is by no means common.

\begin{figure*} % Figure 4
\vspace{0.8cm}
\hspace{-0.2cm}
\centerline{
\scalebox{1.5}{ 
\includegraphics{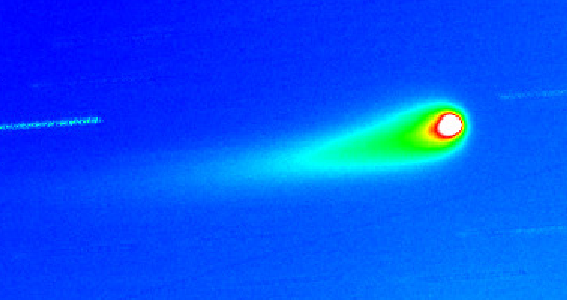}}}

\vspace{0.9cm}                 
\hspace{-0.24cm}
\centerline{
\scalebox{1}{
\includegraphics{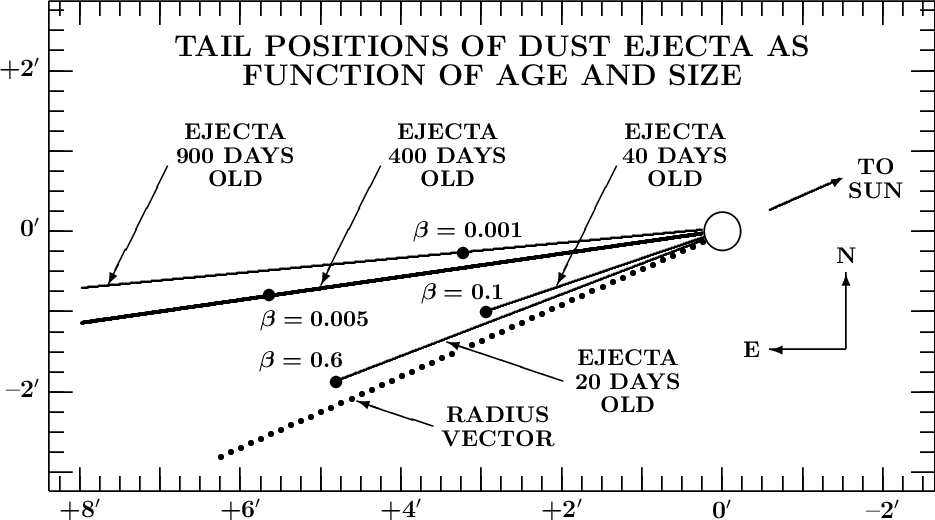}}}
\vspace{0.1cm}
\caption{Dust tail of comet Tsuchinshan-ATLAS.  {\it Upper panel:\/} Cropped
version of the computer processed image of the comet taken by R.\ Naves
with a 30-cm f/9 reflector of the Observatorio Montcabre, Spain, on 2024
June~4.91744 UT, the exposure time of 103~minutes.  North is up, east to the
left.  The image is 12$^\prime\!$.6 on the diagonal.  The tail extends more
than 8$^\prime$ in a position angle of 98$^\circ$.  Note that the tail gets
more narrow as the distance from the head increases, having the appearance
of a teardrop.  {\it Lower panel:\/} Four synchrones that mimick the axial
directions of dust grains ejected from the nucleus (the large open circle)
at different times.  The heavy line, fitting the direction of the observed
tail, shows the ejecta about 400~days old at the time of observation, which
left the nucleus at some 6.6~AU ($\sim$500~days before perihelion); the
other three synchrones (the thinner lines) show the predicted locations of
the tail that would be occupied by dust ejected at 10.6~AU ($\sim$1000~days
before perihelion); at 2.8~AU (155~days before perihelion); and at 2.6~AU
(135~days before perihelion), respectively.  Each synchrone contains
particles of different sizes, the greater ones nearer the nucleus, the
smaller ones at larger distances from it.  The locations of particles of
specific sizes are marked by a dot, the magnitude of the radiation pressure
acceleration $\beta$ (in units of the solar gravitational acceleration)
being indicated.  With $\beta$ essentially varying as an inverse effective
diameter of a particle, the plot suggests that the observed tail consisted
of submillimeter-sized and larger particles that could not be lifted from
the nucleus by sublimating water ice at heliocentric distances greater
than 6~AU.  On the other hand, the image shows quite clearly that the
comet {\it did not\/} eject microscopic dust (submicron- and micron-sized
particles) in any nontrivial amounts in the weeks and months preceding
the time of observation, as the ordinary tail is completely missing.
This anomaly has important implications for the process of fragmentation
that this comet has been subjected to.  (Image courtesy of R.\ Naves,
Observatorio Montcabre, Spain.){\vspace{0cm}}}

\end{figure*}

The key to understanding the length, shape, and orientation of a dust tail
is a computation of the motions of ejected particles relative to the
nucleus as a function of their bulk properties and time of ejection.
The properties (size, shape, bulk density, morphology, and composition)
determine the magnitude of the radiation pressure acceleration that
a particle is subjected to after ejection.  The most widely varying
parameter is the particle size and the acceleration varies inversely
as the size.

To describe the main features of the dust-emission history of comet
Tsuchinshan-ATLAS from Figure~4, one needs to compute a number of
synchrones, loci of particles of different sizes ejected at equal times.
It turns out that the production of dust has been characterized by two
anomalies, one of which is strong emission of submillimeter-sized and
larger particles at heliocentric distances around 6.6~AU, some 500~days
before perihelion.  As the synchrones are crowded, one cannot
determine the effective ejection time from the tail orientation accurately,
but Figure~4 shows no apparent contribution from heliocentric distances
beyond $\sim$10~AU or more than 1000~days before perihelion.  Indeed, the
dust ejected with a zero velocity lined up on 4~June 2024 in a position
angle of 98$^\circ\!$.2 when released 500~days before perihelion (6.56~AU),
in 99$^\circ\!$.5 when 400~days (5.61~AU), in 97$^\circ\!$.3 when 600~days
(7.46~AU), and in 95$^\circ\!$.4 when 1000~days (10.63~AU).  The position
angle is probably measured to slightly better than $\pm$1$^\circ$ in the
image in Figure~4 and the error in ejection times, a little better than
$\pm$100~days, is greater in the direction of early emissions.\footnote{All
position angles are given for equinox J2000.}

As 500~days before perihelion corresponds to mid-May 2023 ($\pm$\,a few months,
because of the uncertainty of position angle measurements), it appears that
the time of major activity (derived from the orientation of the tail)
coincided crudely with the time of discovery of the comet in South Africa
(583~days before perihelion).  Interestingly, according to Yoshida (see
footnote 2){\vspace{-0.04cm}} the comet's light curve was much steeper
{\vspace{-0.05cm}}before 2~March 2023 (${\scriptstyle \propto} \; r^{-14}$)
than after this date (${\scriptstyle \propto} \; r^{-4}$).

The second anomaly is the absence of a tail consisting of freshly ejected
microscopic (submicron- and micron-sized) particles, whose axis is
expected in position angles close to the radius vector.  On 4~June 2024
the radius vector was in position angle 113$^\circ\!$.9, and Figure~4
shows that a 20~days old tail (ejected at 2.56~AU from the Sun) should
be pointing in a position angle of 111$^\circ\!$ and a 40~days old tail
(ejected at 2.84~AU) in 109$^\circ\!$.  Microscopic dust composed of
silicate material would make the tail at least 5$^\prime$ and 18$^\prime$
long, respectively.  If the emissions also contained organic or other
absorbing material, the tail would be about four times longer.  In any
case, it should be long enough to clearly extend beyond the coma
boundaries.

That obviously has not been happening, even though the {\it Af}$\rho$
levels (addressed in Section~2) have not at all been low.  Thus,
the cross-sectional area of solid material in the comet's head is
fairly large, but it does not translate into equivalent amounts
of microscopic dust that would be blown away by radiation pressure into
the tail.  Instead, the fragmentation process appears to essentially
cease when the sizes of fragments reach a certain lower limit.  From
the absence of an ordinary dust tail one cannot say whether the limit
is the size of pebbles or boulders or much bigger objects.  The
relative speeds of these large fragments are low, which may explain
why the central coma of not more than 17\,000~km in radius appears to
be saturated in Naves' image.  In any case, the absence of a tail
containing freshly ejected microscopic dust ejecta is unusual in
a comet that is not depleted in dust.  I will return to this issue
in the next section to address the implications.

In order to verify the conclusions based on the single image in
Figure~4, I have next applied the same method to a set of 31~CCD
tail-orientation observations made, along with his magnitude
observations (Section~2), by A.~Pearce between 13~February and
13~June 2024 (see footnote~4).  Their comparison with the
theoretical loci (synchrones) of dust ejected at 400, 500, and
600~days before perihelion is in Figure~5.  The data between
about 230 and 190~days before perihelion provide striking evidence
in support of the results derived from Figure~4.  Figure~5 also
shows the crowding of the synchrones in June \mbox{(June 4.9 UT
= 114.8~days} before perihelion) and illustrates the exceptional
Sun-comet-Earth geometry that I referred to above.

The issue is as follows:\ dust ejected with a zero velocity and
subjected to a radiation pressure acceleration from the Sun is
always located between the radius vector and the reverse
orbital-velocity vector (i.e., the direction behind the comet).
The angle $\psi$ between the two vectors increases as the comet
proceeds along its path about the Sun; it is near 0$^\circ$
long before perihelion, 90$^\circ$ at perihelion, and approaches
180$^\circ$ long after perihelion.  For a parabolic orbit, an
excellent approximation to the orbit of comet Tsuchinshan-ATLAS,
the angle equals
\begin{equation}
\cos \psi = -{\textstyle \frac{1}{2}} \sqrt{\frac{r}{q}} \sin u
 = \pm\sqrt{1 \!-\! \frac{q}{r}},
\end{equation}
where $r$ and $u$ are, respectively, the heliocentric distance and
true anomaly at the position in the orbit, and $q$ is the perihelion
distance.  The plus sign applies before perihelion, the minus sign
afterwards.

My statement near the beginning of this section that the preperihelion
dust tails of comets are narrow was based on Equation~(11), as $\psi$
provides an upper limit to the tail width.  For example, in a case
relevant to comet Tsuchinshan-ATLAS in April 2024, for \mbox{$q = 0.4$
AU} and \mbox{$r = 3$ AU}, one has \mbox{$\psi = 21^\circ\!$.4}.  In
reality, of course, the tails must be much more narrow than $\psi$. 

The same applies to the tail width in projection onto the plane of
the sky, unless it is distorted because of the position of the Earth
relative to the comet and the Sun.  This happens when the Earth lies
in the comet's orbit plane inside the sector bounded by the two
vectors.  The radius vector and the vector opposite the orbit-velocity
vector then project 180$^\circ$ apart from one another.  When $\psi$
is small, the comet is seen near the Sun in the sky.  But when the
Earth lies inside the extended sector, between the comet and the Sun,
the two vectors also project 180$^\circ$ apart, with the comet near
opposition with the Sun.  And when the Earth is not {\it in\/} but
{\it near\/} the comet's orbit plane, the projected sector gets
only slightly less wide than 180$^\circ$, greatly enhancing the
breadth of any feature inside the sector.  This is what I meant by
the exception.

\begin{figure*}[t] % Figure 5
\vspace{0.18cm}
\hspace{-0.2cm}
\centerline{
\scalebox{0.61}{
\includegraphics{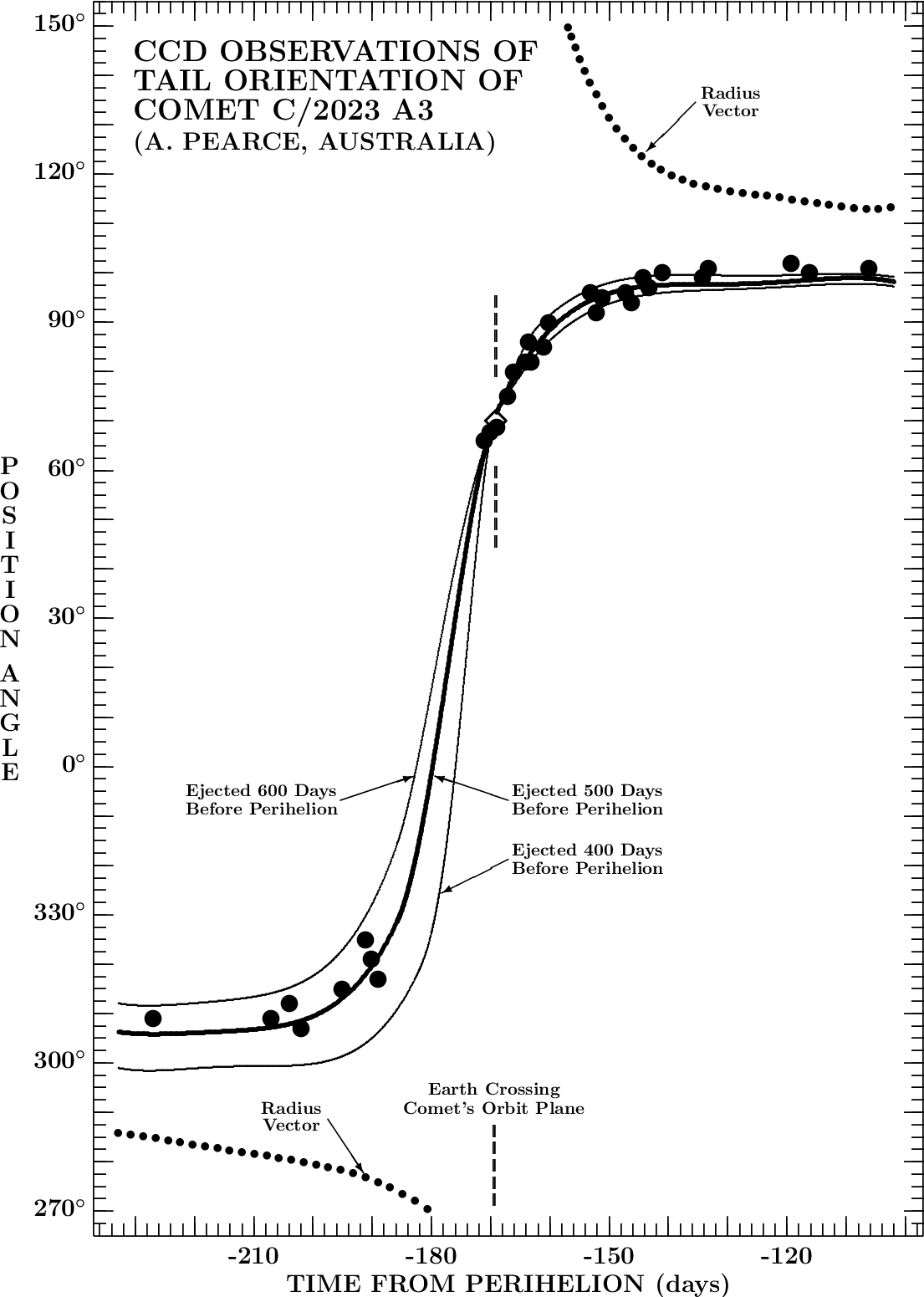}}}
\vspace{-0.05cm}
\caption{Tail orientation of comet Tsuchinshan-ATLAS from CCD
observations by A.\ Pearce, who used his 35-cm f/5
Schmidt-Cassegrain.  The observed position angles (solid
circles; equinox J2000) are compared with the theoretical
synchrones for dust ejected 500~days before perihelion
(thick curve), as well as 100~days earlier and later
(nearly parallel thin lines).  The position angles of
the radius vector are plotted as a dotted curve.  Marked
with a diamond is the time of the Earth's transit across the
comet's orbital plane, 2024 April~11.3~UT or 169.44~days
preperihelion, when the tail pointed at the Sun, in
a position angle of 69$^\circ\!$.7, and technically
became~an~antitail.  The time extends from \mbox{February
4.0 UT = 236.7 days} before perihelion to \mbox{June~22.0 UT =
97.7 days} before perihelion, a total of 139~days.{\vspace{0.6cm}}}
\end{figure*}

As it turned out, comet Tsuchinshan-ATLAS offered the terrestrial observer
exactly this opportunity.  While the {\it true\/} width of the sector grew
very gradually, from $\sim$19$^\circ$ on 10~February to $\sim$21$^\circ$
at the time of the Earth's transit across the orbit plane on 11~April,
the {\it projected\/} width, which already equaled 90$^\circ$ on
10~February, increased to 98$^\circ$ on 1~March, to 124$^\circ$ on
21~March, to 144$^\circ$ on 31~March, and of course to 180$^\circ$
at the transit time.  Subsequently, the projected width dropped to
102$^\circ$ by 20~April, to 48$^\circ$ by 30~April, and to 27$^\circ$
by 20~May, when \mbox{$\psi = 23^\circ$}.  As the projection conditions
continued to worsen, by 9~June the projected width of the sector
diminished to 22$^\circ$, below its true width, \mbox{$\psi = 25^\circ$}.

Because of the strong variations in the projected width of the sector
between the radius vector and the reverse orbital-velocity vector,
Pearce's tail orientation data, accurate to $\pm$2$^\circ$ or so,
contribute unequally to the needs of the applied method.  It turns
out that the degree of angular resolution of dust-emission times (i.e.,
the degree of crowding of synchrones), measured by the width of the
occupied position-angle sector per unit range of relevant dust-emission
times, must be neither too narrow nor too wide to be most useful.
Between 13~February and 22~March, when Pearce made eight tail-orientation
observations, the angular resolution was between about 2$^\circ$ and
4$^\circ$ per month of dust-emission activity.  Then, even though he
observed the comet nine times between 24~March and 8~April, he did
not report the tail a single time.  As Figure~5 shows, the tail was
rotating fast in this period of time (187 through 172~days before
perihelion) and the angular resolution reached 7$^\circ$ per month
of activity.  This temporary ``disappearance'' of the tail indicates
that the emission of dust was continuing over a long period of time and
was projecting as a wide fan, whose surface brightness was obviously
low enough to escape detection.  If instead the dust were emitted in
a brief, powerful outburst, the event would have been detected
regardless of the angular resolution.

On the other hand, following the Earth's crossing the comet's orbit
plane on 11~April, the angular resolution of dust-emission times
remained very low, comparable to 0$^\circ\!$.33 per month of activity
when the image in Figure~4 was taken.  The most useful tail
observations have obviously been those from February and March
(230 through 190~days before perihelion), which show that the
primary emission time was 500--550~days before perihelion, extending
over an estimated period of at least several months.  No observation
by Pearce offers any evidence on a tail made up of freshly ejected
microscopic dust, confirming the conclusion based on the image of
4~June.

The absence of micron-sized and smaller dust grains in the old
emission is unquestionable, but only a very approximate lower limit
on the size of the dust present in the tail could be esimated from
the image in Figure~4.  It was desirable to verify the veracity of
this result.  To do just that, I examined the angular tail lengths
measured by Pearce in the CCD images that he obtained with his
35-cm f/5 Schmidt-Cassegrain.

\begin{figure}[t] % Figure 6
\vspace{0.19cm}
\hspace{-0.18cm}
\centerline{
\scalebox{0.69}{
\includegraphics{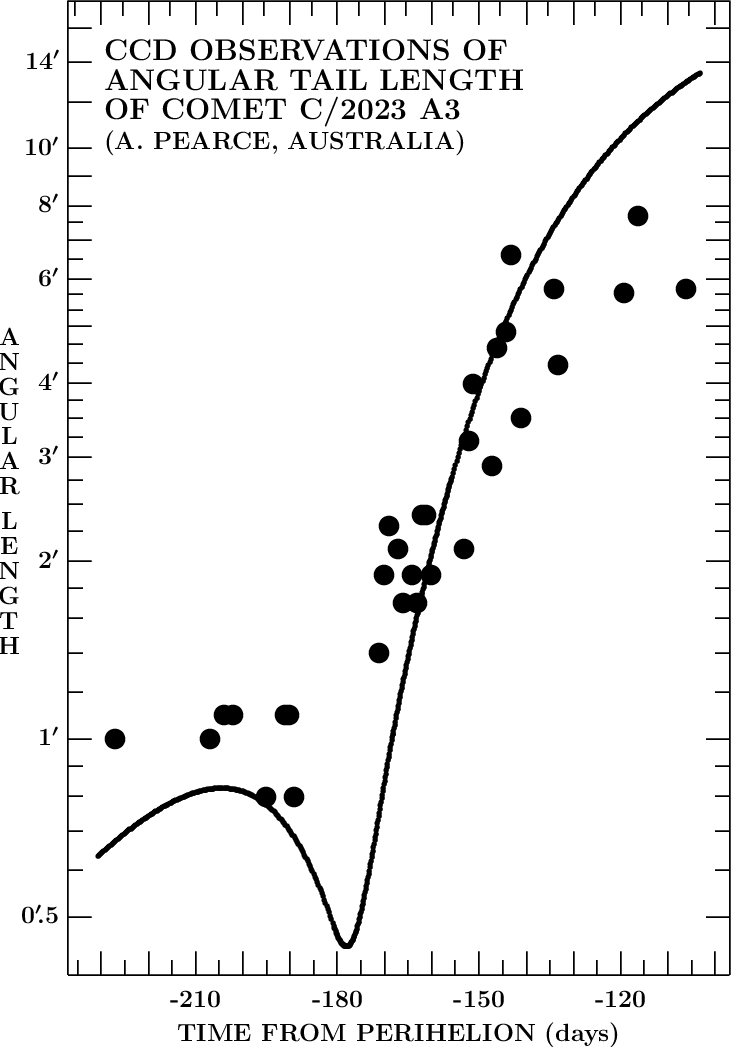}}}
\vspace{-0.05cm}
\caption{Angular tail length of comet Tsuchinshan-ATLAS from
CCD observations by A.\ Pearce, who used his 35-cm f/5
Schmidt-Cassegrain.  Unlike for the tail orientation, the
observations of tail length are affected by observing
conditions and therefore less reliable.  Yet, a fair degree
of agreement is seen between the observations and a predicted
length of a tail consisting of dust particles ejected from
the nucleus 500~days before perihelion and subjected to
radiation pressure accelerations of up to 0.01 the solar
gravitational acceleration.  Such particles are a fraction
of 1~mm across and larger.{\vspace{0.6cm}}}
\end{figure}

The results are presented in Figure 6.  The observations are
compared with the expected angular lengths of a tail consisting
of particles ejected 500~days before perihelion (mid-May 2023;
6.6~AU from the Sun) and subjected to radiation pressure
accelerations not exceeding 0.01~the solar gravitational
acceleration.  At an assumed bulk density of 0.5~g~cm$^{-3}$
such particles are about 0.2~mm or more across.

It is well-known that tail lengths measured in CCD images depend
on the exposure time, observing conditions (such as a degree
of light pollution, air quality, etc.), and other circumstances.
Furthermore, the surface brightness of tails has a tendency to
decline with their age and it is rather common to see the
reported length to get gradually shorter as the surface brightness
drops below the detection threshhold.  Given all these uncertainties,
I deem the fit in Figure~6 rather satisfactory, confirming the
conclusion from Figure~4 that the tail consisted of submillimeter-sized
and larger dust particles.

\section{Fragmentation}
In the preceding sections I have assembled extensive circumstantial
evidence to support a notion that the comet's nucleus is currently,
and has already been for some time, in the process of progressive
fragmentation, which will continue until the point of complete
deactivation and disintegration.  Given the perihelion distance of
0.39~AU, I expect that the object will disappear and cease to exist
as an active comet before perihelion.

Comet Tsuchinshan-ATLAS has been at the risk of perishing before
perihelion by virtue of its arrival from the Oort cloud.  The comets
of this class have a tendency to disintegrate if they are intrinsically
faint and depleted in dust by the time they are near 1~AU from the Sun.
The first condition may be diagnostic of a small, subkilometer-sized
nucleus.  While the dimensions of the nucleus of comet Tsuchinshan-ATLAS
are unknown, the object was neither intrinsically faint nor dust-poor
prior to mid-April.  However, both the observed light curve and
measurements of {\it Af}$\rho$ obtained more recently have suggested
that from mid-April on the comet has been fading and the rate of its
dust production dropping.

In the following I first summarize the evidence presented in the
preceding sections in the context of the proposed progressive
fragmentation scenario.  After that, I briefly address the expected
developments in the near future, before the comet disappears in the
Sun's glare.  Obviously, the results of a search for the comet after
perihelion will ultimately settle the issue of its whereabouts.

The process of fragmentation is proposed to have begun at the latest
around 21~March 2024, 190~days before perihelion and 3.3~AU from the
Sun, when the rate of brightening was abruptly increased from
$r^{-3.5}$ to $r^{-7.2}$.  For an Oort cloud comet such a change is
unusual, because typically the rate of brightening declines with
decreasing heliocentric distance.

The likely cause of the surge or steep brightening (not an outburst!) was
a rapid increase in the cross-sectional area of the sublimating surface
of the nucleus, splitting in quick succession into a large number of
sizable active fragments.  On a short time scale of hours or days the
outgassing fragments were breaking up time and again into smaller
subfragments as the sublimating area continued to grow.  The cloud of
these objects began to slowly expand, running at the same time out of ice.
Its shortage became eventually kind of a firewall that brought the runaway
activity to a stop only 25~days (on 15 April 2024; 165 days before
perihelion and 3~AU from the Sun) after it had started.  As a result,
the light curve reached a peak because the cross-sectional area of
fragments around the nucleus grew no more and their outgassing was
coming to a halt.  The value of {\it Af}$\rho$ started to drop more
rapidly than the brightness, as increasing numbers of fragments
passed the boundary of the zone of 10\,000~km in radius, where this
parameter is measured.  The activity of the comet, whose nucleus
was now reduced in size and much of its ice-rich surface was
gone, began to fade, until the depleted ice was replaced with
fresh supplies from the interior.  In late April, May, and June,
this process appears to have worked haphazardly, on a local scale.
As a result, the comet's brightness was subjected to short-term
fluctuations, which made the comet look as if its activity stalled.
The light curve showed that the comet was struggling, unable to
``pick up steam.''

The observed appearance of the comet as a whole supports the
fragmentation scenario in that starting in late June the
false-color, computer-processed images showed the coma to be
noticeably elongated tailward.  The elongation did not
change appreciably from day to day, suggesting that the flow
of material was steady and/or that the material moved very
slowly, as one would expect of fairly sizable relic objects
left over from the original fragments after all ice sublimated
away.  A stunning piece of evidence that the fragments did not,
on a large scale,~dis\-integrate into microscopic dust is the absence
of an ordinary dust tail adjoining the radius vector.  Significantly,
as of early July, no companion nuclei were detected.

\begin{figure}[b] % Figure 7
\vspace{0.5cm}
\hspace{-0.2cm}
\centerline{
\scalebox{0.6}{
\includegraphics{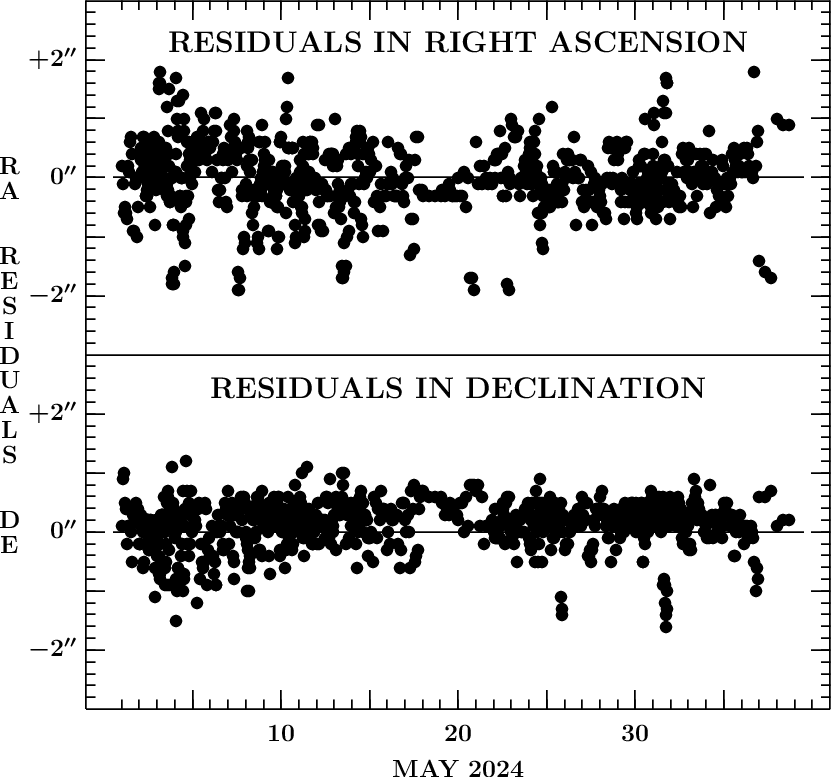}}}
\vspace{-0.03cm}
\caption{Distribution of residuals in right ascension and
declination left by Nakano's recent set of orbital elements of comet
Tsuchinshan-ATLAS (NK 5247) from 860 astrometric observations
made between 1~May and 7~June 2024.  Note the sub-arcsecond
systematic trends in both right ascension and declination.
Nakano rejected all observations whose residuals equaled or 
exceeded $\pm$2$^{\prime\prime\!}$.{\vspace{0cm}}}
\end{figure}

Nakano's sequence of the sets of orbital elements provides evidence
in support of the fragmentation scenario.  The reciprocal original
semimajor axis from his orbital runs that incorporated the astrometric
observations made after the light-curve anomaly is{\vspace{0cm}}
$\sim$30~units of 10$^{-6}$\,AU$^{-1}$ more negative than from the
runs that did not include those observations.  The difference must at
least in part be due to an increased sublimation-driven nongravitational
acceleration on the nucleus that grew smaller because of fragmentation.
The Oort cloud comets moving in orbits with small perihelion
distances have been shown by Marsden et al.\ (1978) to possess a
greater hyperbolic excess than those in orbits with larger
perihelion distances.  These authors presented a formula for the
dependence of $(1/a_{\rm b})_{\rm orig}$ on{\vspace{-0.055cm}} the
perihelion distance, which predicts $-$0.000\,014~AU$^{-1}$ for
comet Tsuchinshan-ATLAS, close to the current value.

Another relevant issue is that of a mean residual, which climbed
from $\pm$0$^{\prime\prime\!}$.32--0$^{\prime\prime\!}$.33 to
$\pm$0$^{\prime\prime\!}$.42 following inclusion of the post-anomaly
data.  This $\sim$30~percent increase in the error means that
either the nuclear condensation became more difficult to measure
accurately (because of its substantially larger dimensions, for
example) or the fit provided by the purely gravitational orbits
was less than entirely satisfactory.  Which of the two reasons
is correct could be decided by carefully inspecting the distribution
of residuals:\ in the former case it should be broader than before
but still essentially random, while one should see systematic
trends in the latter case.

\begin{table}[b] % Table 3
\vspace{0.5cm}
\hspace{-0.2cm}
\centerline{
\scalebox{1}{
\includegraphics{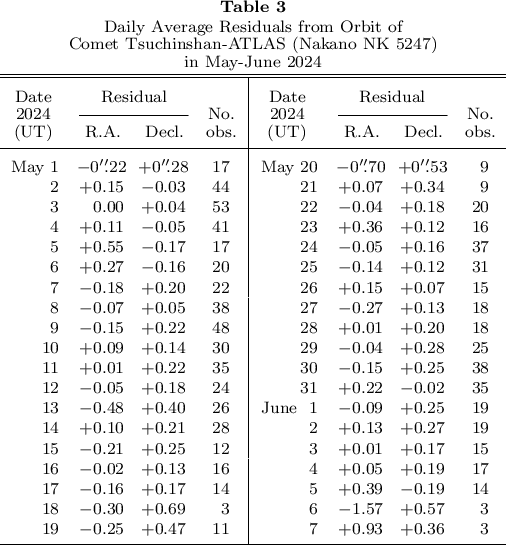}}}
\vspace{0cm}
\end{table}

Because of time constraints, I have inspected the problem rather
superficially.  I have focused on the residuals from Nakano's recent set
of orbital elements (NK 5247) left by 860 astrometric observations
(a small fraction of more than 4000 data points used) made between
1~May and 7~June 2024.  The results of my inspection are apparent from
Figure~7 and Table~3, which consistently show sub-arcsecond deviations
from a random distribution.

The distribution of residuals in right ascension in the upper half
of the figure displays a wide letter V:\ most residuals in early May
are positive trending down toward negative numbers, in late May
and June back up toward positive ones.  Four of the first six
tabulated daily averages are positive and only one is negative.
In the period of 15--20 May all averages are negative, while five of
the last six averages are positive.  In declination, the systematic
trend in the figure is immediately obvious, as a great majority of
the residuals is positive.  The table shows that only six out of the
38~daily averages are negative.  Whether or not these effects are
products of a nongravitational acceleration remains to be seen.

The only measurement of the water production rate that I am aware
of at this time implies a sublimation area of 15~km$^2$, a rather
bewildering piece of information.  By itself it allows two very
different interpretations, neither one of which looks reasonable
to me:\ an enormous nucleus or a nucleus in a highly advanced
stage of fragmentation.

In the short-period of time, over which the comet can still be
observed from some locations, not much additional information
can be gathered over what is currently known.  It is hoped that
the existing databases will further be expanded:\ the light curve,
the curve of {\it Af}$\rho$, the orbital arc, and more data
on the water production rate.  Of special interest is evidence
of the continuing absence of a tail made up of recent micron-
and submicron-sized ejecta in images of the longest possible
exposure.

I expect that the light and {\it Af}$\rho$ curves will continue
to drop in an uneven fashion, even though the absolute magnitude
of 11.8 from the data in Figure~2 may be too pessimistic.  Whether
the comet will undergo another surge of brightness of the kind
experienced in late March and early April depends on the size of
the nucleus; a chance is that there will be none.  Essentially
the same uncertainty applies to {\it Af}$\rho$, whose variations
with time appear to be governed by the same processes.

In the field of orbit determination, I predict that the mean
residual from {\it gravitational\/} orbital solutions will
continue to grow and the reciprocal original semimajor axis
will continue to get more hyperbolic.  Sooner or later it will
be necessary to introduce the nongravitational terms.  Such
orbital solutions should be attempted earlier rather than
later and tested by how much the mean residual has been
improved.

New data for the water production rate will be extremely useful
not only for getting a more integrated picture of the comet's
physical evolution, but also for verifying the rate from the
end of May.

One dataset that unfortunately will not become~available is
a close-up view of the comet's nuclear region.  One can only
envision an image looking like superimposed multiple
exposures of the nuclear region of comet C/2019~Y4, taken with
the Hubble Space Telescope's camera (Ye et al.\ 2021) --- a field
with an enormous cloud of barely glimmering fragments, some brighter,
some fainter.  In the final phase of fragmentation, {\small \bf
increasing numbers of devolatilized, fractured refractory solids
stay assembled in dark, highly porous, and exotically shaped blobs}
that eventually become undetectable as they gradually disperse in
space.  One can only speculate that some (the largest?) of these
blobs could possibly look like 1I/`Oumuamua.\\[0.2cm]

\section{Conclusions}
The purpose of this paper is not to disappoint comet observers
who have been looking forward to a new naked-eye object this
coming October, but to present scientific arguments that do not
appear to substantiate such hopes.  Even though the prognostication
of preperihelion disintegration of comets is admittedly a very
risky undertaking, I believe that the time has come to go ahead
with it.

Comet Tsuchinshan-ATLAS exhibits traits that are diagnostic of
extensive fragmentation of the nucleus, even though no distinct
companion has been observed as of early July.  Some of the
features resemble the performance of past members of the Oort
cloud that disintegrated before reaching perihelion at a
heliocentric distance substantially smaller than 1~AU.  It is
primarily the peculiar light curve, as pointed out by Ferrin,
as well as the parallel changes in {\it Af}$\rho$, especially
a surge in late March through early April, a sharp peak in
mid-April, when the comet was at 3~AU from the Sun, and the
subsequent fading with superimposed fluctuations, which are
likely to testify to additional events of local fragmentation.
Also worrisome is the apparent tailward elongation of the coma
seen in June.  The extremely high water production rate needs
confirmation.

Independent support is provided by two effects in the comet's
orbital motion:\ (i)~a hyperbolic{\vspace{-0.03cm}} shift of
30~units of 10$^{-6}$\,AU$^{-1}$ in the reciprocal original semimajor
axis after inclusion of astrometric observations made following
the light-curve anomaly and (ii)~parallel increase in the mean
residual from $\pm$0$^{\prime\prime\!}$.32--0$^{\prime\prime\!}$.33
to $\pm$0$^{\prime\prime\!}$.42.  The culprit appears to be a
sublimation-driven nongravitational acceleration on the nucleus
whose dimensions have been rapidly diminishing because of its
continuing fragmentation.  Even though detected systematic trends
have as yet been in a sub-arcsecond range, they soon may grow and
require the incorporation of the nongravitational terms in
orbit determination efforts.

Most unusual is the continuing absence of an ordinary dust tail,
which means that large amounts of dry, fractured solid material
do not disintegrate into microscopic dust, but stay assembled in
dark and highly porous bizarre bodies that I refer to above as
{\it blobs\/}.  Once they disperse in space,\,they are nearly
impossible to detect,\,yet they may be omnipresent though perhaps
short lived.

{\sf Added after final editing:} As observations continue~and the data
in the paper get increasingly incomplete, some of the numerical results
may be affected.  Apologies. \\[-0.32cm]
\begin{center}
{\footnotesize REFERENCES}
\end{center}

\vspace{-0.45cm}
\begin{description}
{\footnotesize
\item[\hspace{-0.3cm}]
A$\!$'Hearn,\,M.\,F.,~Schleicher,\,D.\,G.,~Millis,\,R.\,L.,\,et\,al.\,1984,\,Astron.\,J.,{\linebreak}
 {\hspace*{-0.6cm}}89, 579
\\[-0.57cm]
\item[\hspace{-0.3cm}]
Ahuja, G., Aravind, K., Sahu, D., et al.\ 2024, Astron.\ Tel., 16637\,\,\,\,\,
\\[-0.57cm] % carbon depleted
\item[\hspace{-0.3cm}]
Belton, M.\ J.\ S.\ 1965, Astron.\ J., 70, 451
\\[-0.57cm]
\item[\hspace{-0.3cm}]
Bortle, J.\ E.\ 1991, Int.\ Comet Quart., 13, 89
\\[-0.57cm]
\item[\hspace{-0.3cm}]
Brandt, J.\ C.\ 1961, Astrophys.\ J., 133, 1091
\\[-0.57cm]
\item[\hspace{-0.3cm}]
Green, D.\ W.\ E., ed.\ 2023, Cent.\ Bur.\ Electr.\ Tel.\ 5228
\\[-0.57cm]
\item[\hspace{-0.3cm}]
Green, D.\ W.\ E., ed.\ 2024, Cent.\ Bur.\ Electr.\ Tel.\ 5404
\\[-0.57cm]
\item[\hspace{-0.3cm}]
Jewitt, D., \& Luu, J.\ 2019, Astrophys.\ J., 883, L28 (6pp)
% C/2019 J2 dead at 1.9 AU
\\[-0.57cm]
\item[\hspace{-0.3cm}]
Kr\'olikowska, M., \& Dones, L.\ 2023, Astron.\ Astrophys., 678, A113
\\[-0.57cm]
\item[\hspace{-0.3cm}]
Marsden,\,B.\,G., Sekanina,\,Z., \& Everhart,\,E.\,1978, Astron.\,J.,\,78,\,64
\\[-0.57cm]
\item[\hspace{-0.3cm}]
Meech,\,K.\,J.,\,Pittichov\'a,\,J.,\,Bar-Nun,\,A.,\,et\,al.\,2009, Icarus,\,201,\,719
% {\linebreak}
% {\hspace*{-0.6cm}}719
\\[-0.57cm]\
%
% \item[\hspace{-0.3cm}]
% Nakano, S.\ 2001, Publ.\,Astron.\,Soc.\,Japan, 53, 931
% \\[-0.57cm]
%
\item[\hspace{-0.3cm}]
Osterbrock, D.\ E.\ 1958, Astrophys.\ J., 128, 95
\\[-0.57cm]
%
% \item[\hspace{-0.3cm}]
% Pingr\'e,\,A.\,G.\,1783, Com\'etographie ou\,Trait\'e historique et\,th\'eorique{\linebreak}
% {\hspace*{-0.6cm}}des com\`etes. Tome Premier. Paris: Imprimerie Royale
% \\[-0.57cm]
%
\item[\hspace{-0.3cm}]
Roemer, E.\ 1962, Publ.\ Astron.\ Soc.\ Pacific, 74, 351
\\[-0.57cm]
\item[\hspace{-0.3cm}]
Sekanina, Z.\ 1973, Astrophys.\ Lett., 14, 175
\\[-0.57cm]
\item[\hspace{-0.3cm}]
Sekanina, Z. 2019, eprint arXiv:1903.06300
\\[-0.64cm]
\item[\hspace{-0.3cm}]
Ye,\,Q., Jewitt,\,D., Hui,\,M.-T., et al.\,2021, Astron.\,J.,\,162,\,70\,(13pp)}
\vspace{-0.34cm}

\end{description}
\end{document}